\documentclass{article}

\usepackage[preprint]{neurips_2025}

\usepackage[utf8]{inputenc} 
\usepackage[T1]{fontenc}    
\usepackage{hyperref}       
\usepackage{url}            
\usepackage{booktabs}       
\usepackage{amsfonts}       
\usepackage{nicefrac}       
\usepackage{microtype}      
\usepackage{xcolor}         

\usepackage{graphicx}
\usepackage{adjustbox}
\usepackage{makecell}
\usepackage{subcaption}
\title{NOVA: A Benchmark for Anomaly Localization and Clinical Reasoning in Brain MRI}

\author{%
\textbf{Cosmin I. Bercea}$^{1,3}$ \quad 
\textbf{Jun Li}$^1$ \quad
\textbf{Philipp Raffler}$^{2}$ \quad 
\textbf{Evamaria O. Riedel}$^{2}$ \quad 
\textbf{Lena Schmitzer}$^{2}$ \\
\textbf{Angela Kurz}$^{2}$ \quad 
\textbf{Felix Bitzer}$^{2}$ \quad 
\textbf{Paula Roßmüller}$^{2}$ \quad
\textbf{Julian Canisius}$^{2}$ \quad 
\textbf{Mirjam L. Beyrle}$^{2}$ \\ 
\textbf{Che Liu}$^5$ \quad
\textbf{Wenjia Bai}$^5$ \quad
\textbf{Bernhard Kainz}$^{4,5}$ \quad 
\textbf{Julia A. Schnabel}$^{1,3,6}$ \quad 
\textbf{Benedikt Wiestler}$^{1,2}$ 
\\
$^1$Technical University of Munich \quad
$^2$Klinikum Rechts der Isar \quad
$^3$Helmholtz Center Munich \\
$^4$FAU Erlangen-Nürnberg \quad
$^5$Imperial College London \quad
$^6$King's College London \\
\texttt{cosmin.bercea@tum.de}
}

\begin{document}

\maketitle

\begin{abstract}
In many real-world applications, deployed models encounter inputs that differ from the data seen during training. Out-of-distribution detection identifies whether an input stems from an unseen distribution, while open-world recognition flags such inputs to ensure the system remains robust as ever-emerging, previously \emph{unknown} categories appear and must be addressed without retraining.
Foundation and vision-language models are pre-trained on large and diverse datasets with the expectation of broad generalization across domains, including medical imaging.
However, benchmarking these models on test sets with only a few common outlier types silently collapses the evaluation back to a closed-set problem, masking failures on rare or truly novel conditions encountered in clinical use.

We therefore present \emph{NOVA}, a challenging, real-life \emph{evaluation-only} benchmark of $\sim$900 brain MRI scans that span 281 rare pathologies and heterogeneous acquisition protocols. Each case includes rich clinical narratives and double-blinded expert bounding-box annotations. Together, these enable joint assessment of anomaly localisation, visual captioning, and diagnostic reasoning. 
Because NOVA is never used for training, it serves as an \textit{extreme} stress-test of out-of-distribution generalisation: models must bridge a distribution gap both in sample appearance and in semantic space.  
Baseline results with leading vision-language models (GPT-4o, Gemini 2.0 Flash, and Qwen2.5-VL-72B) reveal substantial performance drops across all tasks, establishing NOVA as a rigorous testbed for advancing models that can detect, localize, and reason about truly unknown anomalies.
\end{abstract}

\section{Introduction}

Generalization under distribution shift remains a central unsolved challenge in machine learning~\cite{feng2022promptdet,zhu2024survey}. Despite advances in large-scale pretraining and transfer learning~\cite{dosovitskiy2020image,radford2021learning}, most models fail to reliably detect or reason about previously unseen categories or domains at test time. Anomaly detection, the task of identifying deviations from a given normative distribution, \emph{e.g.}, samples exclusively from healthy patients, represents an extreme stress-test of out-of-distribution (OOD) generalization due to the open-ended and unpredictable nature of anomalies. While OOD generalization has been extensively studied in natural image classification~\cite{hendrycks2018deep, recht2019imagenet, koh2021wilds}, it remains underexplored in healthcare. Medical data presents extreme heterogeneity, rare event frequencies, and non-standardized acquisition protocols, making it a worst-case scenario for evaluating model robustness to distribution shift. Detecting anomalies—potential pathologies—in imaging is often the first and most challenging step of the diagnostic process. Providing effective assistance to physicians at this stage has the potential to substantially improve clinical outcomes.

In medical imaging, unsupervised anomaly detection (UAD) methods~\cite{ruff2021unifying,  bercea2025evaluating} are trained exclusively on healthy anatomy and identify deviations from this learned distribution as potential pathologies. 
While recent methods demonstrate strong performance on curated benchmarks~\cite{zimmerer2019unsupervised, schlegl2019f, chen2020unsupervised, wolleb2022diffusion, schluter2022natural, bercea2024diffusion}, they remain insufficiently reliable in the wild, particularly in high-stakes settings like clinical triage and health screening, where specificity and robustness to rare clinical presentations are essential~\cite{bercea2023generalizing, emergency}.
This challenge is particularly acute in magnetic resonance imaging (MRI) of the brain, where radiologists must detect subtle and diverse abnormalities across patient populations and heterogeneous imaging protocols.

The fundamental bottleneck lies in the datasets used for validation. Most existing benchmarks define anomalies through fixed categories, inducing implicit data leakage: although models are trained on healthy data, test sets remain constrained to known abnormality types. This narrows the evaluation to familiar distributions and undermines open-set detection. Datasets such as BraTS~\cite{menze2015brats}, ATLAS~\cite{atlas}, and ISLES~\cite{isles} were designed for segmentation and primarily capture canonical disease patterns, causing model development to converge on closed-set optimization rather than true discovery of unknown conditions.

The medical out-of-distribution analysis challenge (MOOD)~\cite{zimmerer2022mood} introduced synthetic anomalies to simulate unknown deviations. However, \emph{real} anomalies from rare or previously unobserved diseases remain essential for clinical relevance. fastMRI+~\cite{zhao2022fastmri+} provided incremental pathology variability through bounding box annotations of brain and knee MRI scans~\cite{zbontar2018fastmri}, yet lacked the pathology heterogeneity and structured clinical metadata necessary to reflect clinical variability.

Detecting an abnormality alone does not satisfy clinical requirements. Radiologists must localize pathologically suspicious regions, assess severity, distinguish them from imaging artefacts, and formulate a differential diagnosis based on patient history and imaging findings. No existing dataset reflects this full diagnostic workflow, limiting prior benchmarks to binary detection and systematically failing to capture clinically meaningful information in generated text or diagnostic predictions~\cite{mahmood2025benchmarking}. Data sharing restrictions and poor standardization have further constrained the development of vision-language models (VLMs) in medicine.

\begin{figure}[tb!]
\centering
\includegraphics[width=\linewidth]{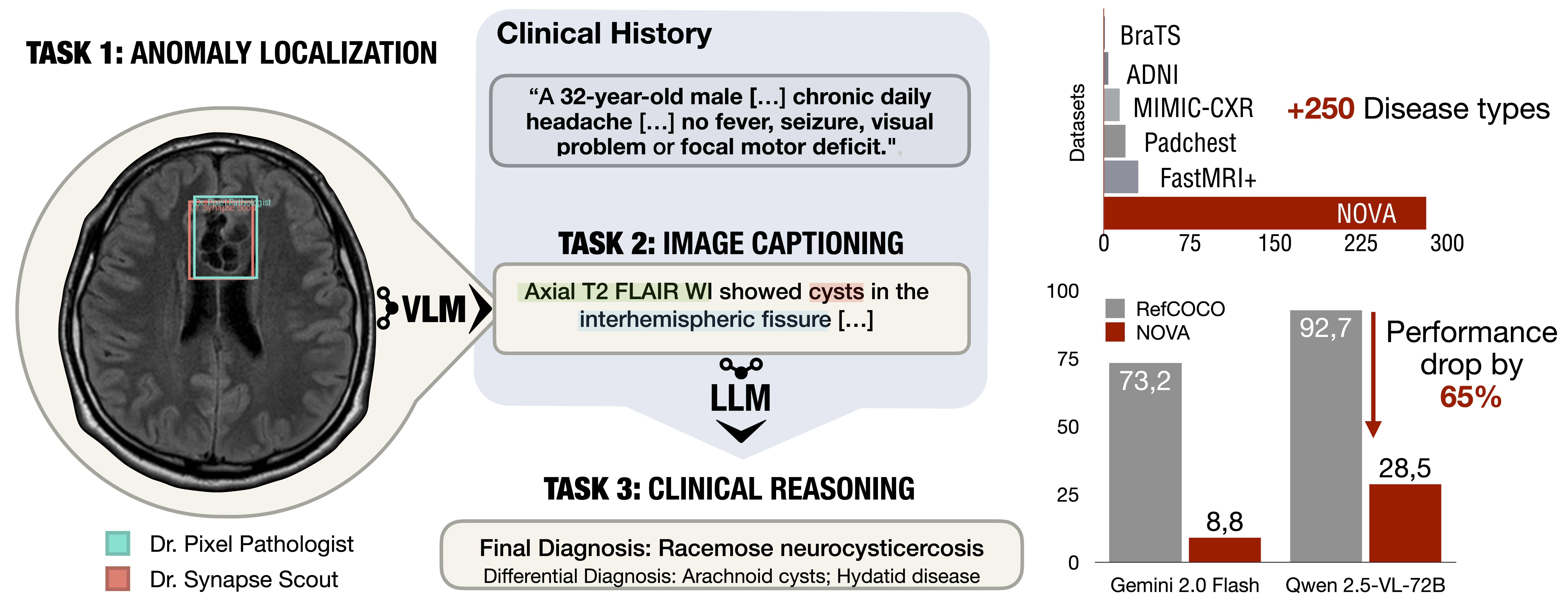}
\caption{Overview of the NOVA benchmark. Task 1: Anomaly localization: models predict bounding boxes identifying abnormal regions in brain MRI; ground truth annotations from two independent radiologists are shown. Task 2: Image captioning: models generate a brief diagnostic description from the MRI image. Task 3: Diagnostic reasoning: models predict the final diagnosis by integrating clinical history and image findings. NOVA establishes the first benchmark designed to systematically evaluate vision-language models (VLMs) and large language models (LLMs) for rare anomaly localization, clinical description, and multimodal diagnostic reasoning in brain MRI.}
\label{fig::teaser}
\end{figure}

Therefore, NOVA establishes a new benchmark for evaluating, detecting, and reasoning on unexpected abnormalities in clinical brain MRI, as illustrated in Figure~\ref{fig::teaser}. The dataset comprises 906 brain MRI scans spanning 281 rare and diagnostically diverse pathologies from Eurorad, enriched with detailed clinical narratives. Each case is independently annotated by at least two radiologists with bounding boxes identifying suspected abnormalities. NOVA uniquely enables joint evaluation of anomaly localization, visual captioning, and diagnostic reasoning under real-world clinical heterogeneity. It is explicitly designed as an evaluation-only benchmark to serve as an extreme stress-test of OOD generalization, requiring models to bridge distribution shifts in both visual and semantic space.

We benchmark state-of-the-art vision-language models, including GPT-4o, Gemini 2.0 Flash, and Qwen2.5-VL-72B, on NOVA. Results reveal substantial performance degradation across all tasks, underscoring the urgent need for benchmarks that reflect the demands of open-world clinical reasoning.
\section{Related Work}

Anomaly detection, OOD detection, and novelty detection have received sustained focus in computer vision and machine learning, with advances across tasks from natural image understanding to industrial inspection~\cite{pidhorskyi2018generative, li2021cutpaste, ood1, ood2, ood3, roth2022towards, wang2024bridging}.
Despite this progress, transferring these methods to medical imaging remains challenging. 
The concept of normality in medicine is inherently ambiguous, varying across individuals, imaging protocols, and institutions. 

Clinical anomalies are often rare and highly heterogeneous, making them ill-suited for evaluation protocols that treat a selected set of predefined categories as representative out-of-distribution cases. 
The distinction between healthy and abnormal tissue is frequently subtle and localized, with considerable overlap between in-distribution and out-of-distribution regions within the same image. As a result, approaches that excel on constrained datasets such as MVTec-AD~\cite{bergmann2019mvtec} fail systematically under the extreme clinical variability of real-world neuroimaging~\cite{hong2024out}.

In medical imaging, unsupervised anomaly detection models learn the normative distribution of healthy anatomy to identify deviations as candidate pathologies~\cite{bercea2025evaluating}. Large healthy population datasets, including IXI~\cite{ixi}, CamCAN~\cite{camcan}, and UK Biobank~\cite{ukbiobank} are valuable for normative modeling and population studies, but they are ill-suited for evaluating anomaly detection, as they lack pathological cases and localized anomaly annotations. 
Datasets such as ADNI~\cite{adni} and OASIS~\cite{oasis} focus exclusively on Alzheimer’s disease and neurodegeneration, providing only narrow coverage of pathologies. Similarly, condition-specific datasets, including MSLUB~\cite{mslub} for multiple sclerosis lesions, ATLAS~\cite{atlas} and ISLES~\cite{isles} for Stroke lesions, and BraTS~\cite{menze2015brats} for brain tumors, support segmentation of predefined abnormalities but offer no framework for open-set detection or evaluation of vision-language reasoning in clinical contexts.

In parallel, large-scale vision-language datasets in medical imaging focus exclusively on chest radiographs. MIMIC-CXR~\cite{mimic} and PadChest~\cite{padchest, padchest-gr} integrate images with radiology reports for multimodal learning but are entirely disconnected from brain MRI. Hamamci et al.~\cite{hamamci2024developing} introduced the CLM3D dataset and corresponding VLM3D challenge for developing generalist vision-language models in 3D medical imaging. However, CLM3D targets thoracic CT and focuses on common abnormality classification, report generation, and text-conditioned image synthesis, without addressing rare disease detection, anomaly localization, or open-world clinical reasoning. 

Despite the critical technical and clinical need, a comprehensive neuroimaging benchmark remains absent. Brain MRI analysis presents significant technical challenges due to the wide spectrum of pathologies and their diverse appearances, ranging from localized lesions to diffuse structural alterations, coupled with inherent technical variability. Clinically, a clear need exists as most rare diseases are neurological or have neurological manifestations~\cite{brain}, positioning brain MRI centrally in patient care.
NOVA establishes the first rigorous benchmark for systematically evaluating these capabilities under the real-world variability and diagnostic uncertainty of clinical brain MRI.
\section{Dataset Description}

\begin{figure}[tb!]
\centering
\includegraphics[width=\linewidth]{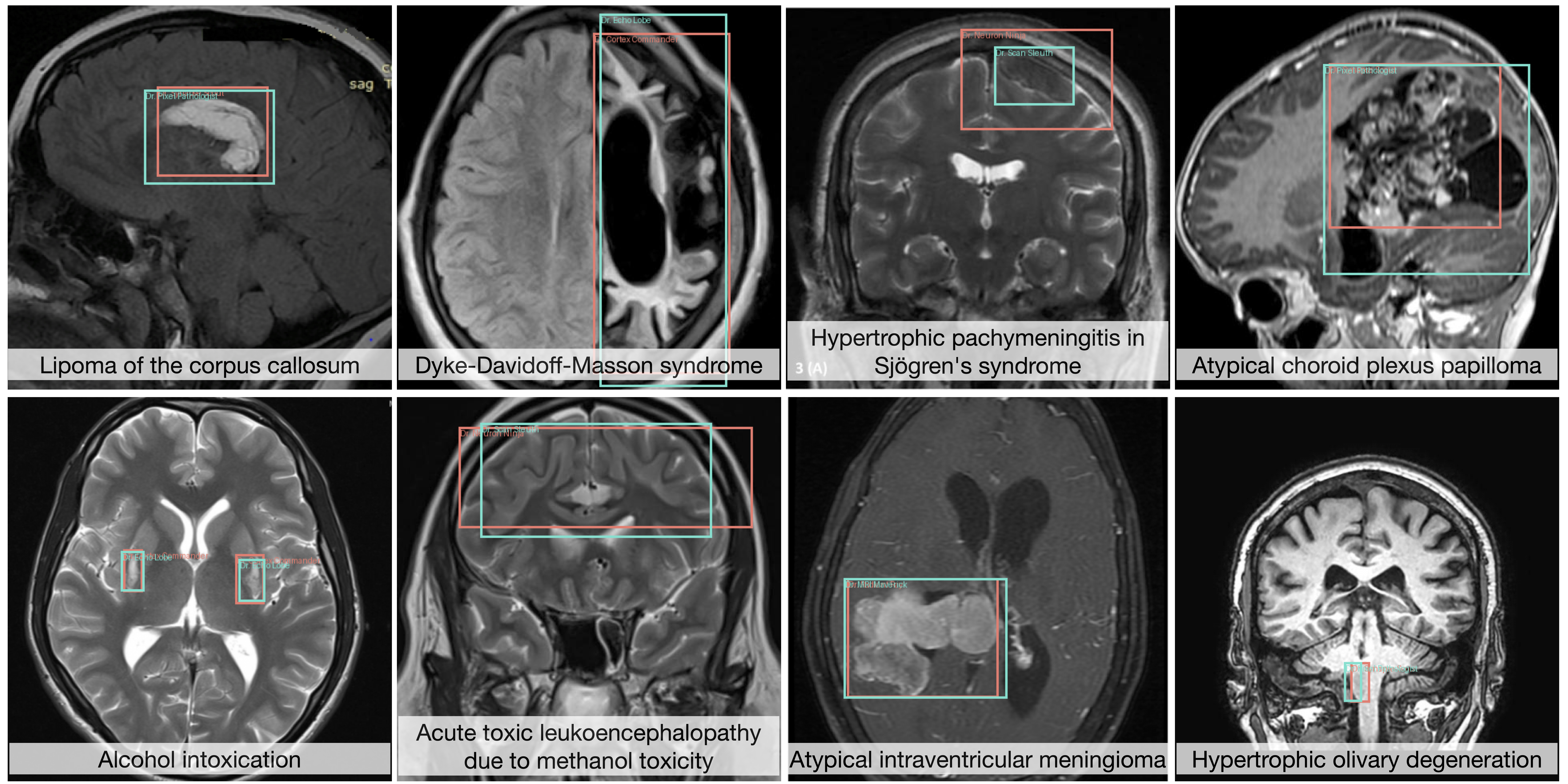}
\caption{Representative brain MRI scans from the NOVA dataset illustrating the diversity of anatomical planes, MRI sequences, and pathological conditions. Radiologist-provided bounding box annotations are overlaid. The examples include rare congenital malformations, toxic and metabolic encephalopathies, and inflammatory or neoplastic lesions—capturing the broad radiological spectrum.}
\label{fig::data_overview}
\end{figure}

We curated the NOVA dataset to establish an evaluation benchmark for vision-language model generalization under extreme clinical variability in brain MRI. We sourced cases from Eurorad, a peer-reviewed educational platform operating under a Creative Commons Attribution-NonCommercial-ShareAlike 4.0 International License\footnote{\url{https://www.eurorad.org/node/38655}}. To comply with licensing requirements, we included only cases published after July 6, 2015. We filtered the dataset to include all cases from the ``Neuroradiology'' category and manually excluded non-relevant content such as CT, spine MRI, clinical photographs, and other non-MRI data. This ensured consistent imaging modality and anatomical focus.

We collected a total of 906 brain MRI scans spanning 281 unique diagnoses. We retained all images in their original form without preprocessing, cropping, or normalization to preserve the full clinical variability essential for evaluation. We preserved the naturally imbalanced long-tailed distribution of rare diseases to reflect real-world case frequencies. Representative examples illustrating the diversity of imaging planes, sequences, and pathologies are shown in Figure~\ref{fig::data_overview}.

\subsection{Dataset Composition}

\begin{figure}[tb!]
\centering
\includegraphics[width=\linewidth]{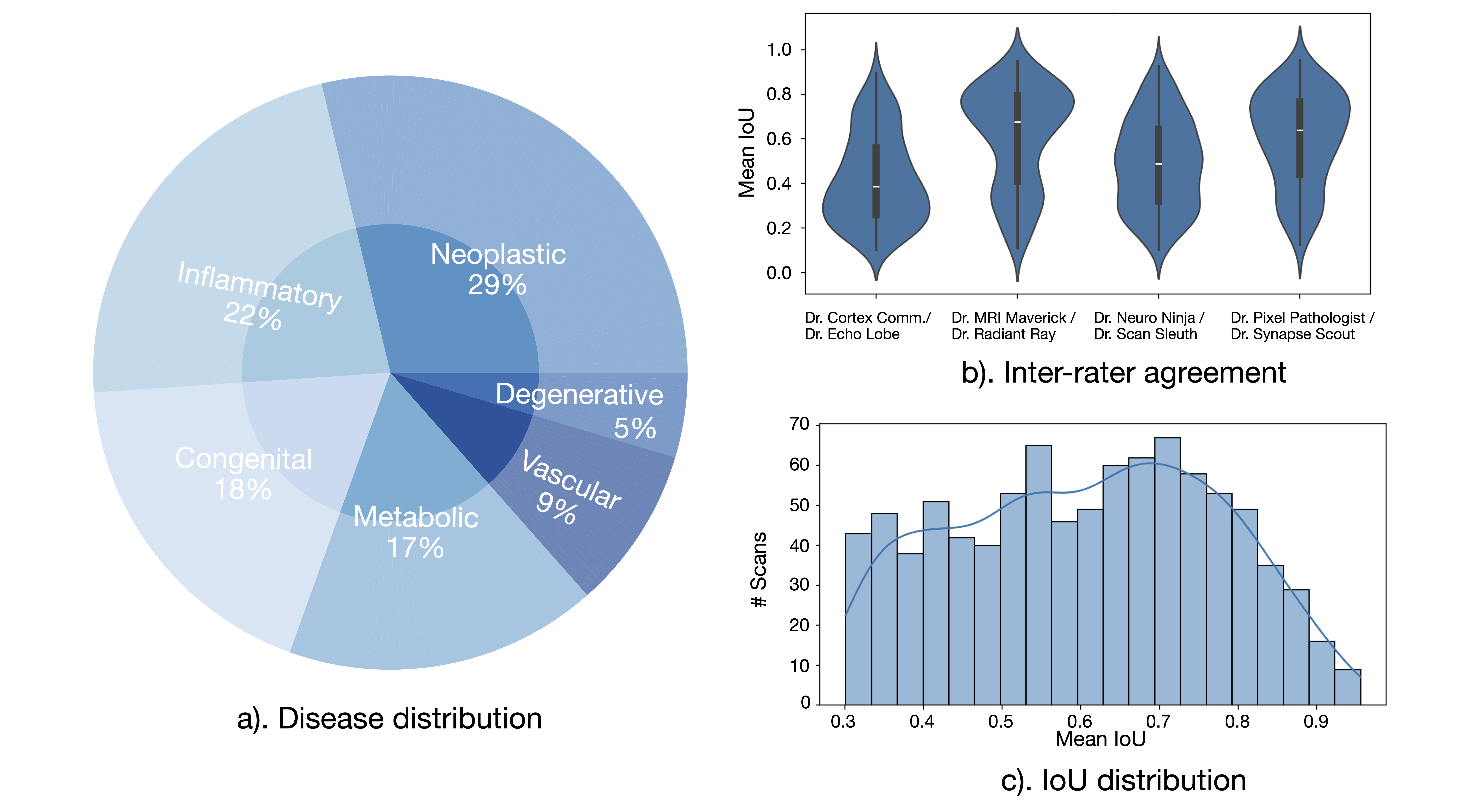}
\caption{Dataset composition and annotation quality in NOVA. (a) Distribution of cases across six diagnostic categories. (b) Inter-rater agreement as mean intersection over union (IoU) between radiologist pairs. (c) Histogram of IoU scores across all scans.}
\label{fig::data_stats}
\end{figure}

NOVA captures the diagnostic heterogeneity of clinical brain MRI. We included axial, sagittal, and coronal planes across standard sequences, including T1-weighted, T2-weighted, and FLAIR imaging. We manually grouped cases into six diagnostic categories: neoplastic, neurodegenerative, inflammatory, congenital, metabolic, and vascular pathologies. Figure~\ref{fig::data_stats}a) shows the long-tailed distribution of diseases and the rarity of many conditions, which present unique challenges for model evaluation. The dataset's 281 unique diagnosis labels exceed the diversity of existing brain MRI benchmarks by an order of magnitude. We summarize additional statistics on patient demographics, sequence types, and imaging parameters in Appendix.

\subsection{Annotation Process and Quality Control}

We implemented a rigorous multi-stage protocol to obtain high-quality anomaly localization annotations. Eight neuroradiology residents annotated the dataset using a custom web-based platform (Appendix). Each case was independently labeled by two readers, who reviewed the full original Eurorad clinical description and associated metadata to inform their annotations.

Inter-rater agreement was computed using a greedy matching algorithm that maximized intersection over union (IoU) between boxes. Annotations with IoU > 0.3 were merged into consensus labels. For 247 cases with persistent disagreement, a senior board-certified neuroradiologist (15+ years experience) adjudicated the final ground truth.

Figure~\ref{fig::data_stats}b shows the distribution of inter-reader agreement, illustrating the inherent variability in clinical anomaly localization. Figure~\ref{fig::data_stats}c presents the overall IoU distribution across annotations, highlighting the diagnostic ambiguity and challenges of consistently identifying anomalies in real-world clinical imaging.

\subsection{Data Format and Benchmark Design}

We release all brain MRI scans as uniformly sized 480$\times$480 grayscale PNG slices. We provide accompanying clinical metadata, including clinical history, patient demographics, imaging information, radiologist image captions, and bounding boxes for detected abnormalities in csv files.

We explicitly designed NOVA as an evaluation-only dataset. Each case represents a unique diagnosis, and we do not provide predefined train, validation, or test splits. This enforces a zero-shot evaluation setting for all models, requiring them to generalize to previously unseen cases.

We publicly release NOVA on Hugging Face Datasets\footnote{\url{hhttps://huggingface.co/datasets/Ano-2090/Nova}} under the same Creative Commons Attribution-NonCommercial-ShareAlike 4.0 license as the Eurorad source. The dataset is distributed solely for non-commercial research to enable reproducible evaluation of vision-language models under realistic clinical conditions.
\section{Benchmark Tasks}

NOVA defines a comprehensive evaluation suite to assess the capabilities of vision-language models under realistic and clinically relevant conditions. These three tasks defined here reflect the sequential decision-making process of radiologists, progressing from anomaly localization to image description and diagnostic interpretation, enabling a realistic benchmarking.

\subsection{Task 1: Anomaly Localization}

This task requires models to detect and localize abnormalities within brain MRI scans, regardless of the patient’s eventual diagnosis. Clinically, this is a relevant task, as most medical errors actually stem from not seeing a pathology at all~\cite{Kim2014Fool}. Models must predict one or more bounding boxes per image corresponding to abnormal regions, using radiologist-annotated ground truth as reference. Performance is measured using mean average precision at intersection over union thresholds of 0.3 (mAP@30), 0.5 (mAP@50), and the COCO-style averaged mAP across thresholds from 0.50 to 0.95 in 0.05 increments (mAP@[50:95]). The benchmark also reports the number of correctly detected versus missed pathologies per case to reflect the clinical priority of minimizing false negatives. The dataset includes cases with multiple annotated abnormalities, providing a uniquely difficult evaluation setting for object detection under open-world conditions and rare disease variability.

\subsection{Task 2: Image Description}

This task measures the ability of models to generate clinically meaningful descriptions of brain MRI scans, an important prerequisite for making the correct diagnosis and in clinical communication. Each image is paired with an expert-generated caption describing the imaging findings. Evaluation uses case-insensitive exact keyword matching to compute precision, recall, and F1-score across the full keyword set. Modality-specific terms (such as \textit{flair}, \textit{axial}, \textit{sagittal}, \textit{t1}, \textit{t2}, \textit{coronal}, \textit{dwi}, \textit{t1w}, \textit{t2w}, \textit{weighted}) are evaluated separately from non-modality keywords capturing clinical content. Binary classification accuracy and F1-score for normal versus abnormal classification are also reported. Sentence-level generation quality is evaluated using BLEU~\cite{papineni-etal-2002-bleu} and METEOR~\cite{banerjee-lavie-2005-meteor}.

\subsection{Task 3: Diagnostic Reasoning}

This task tests whether models can integrate clinical context and imaging observations to predict a diagnosis, arguably the "supreme discipline" in medical decision-making. Each case provides a brief clinical history and corresponding image caption as input, and the model must generate a free-text diagnostic label. Performance is reported as Top-1 accuracy (exact match with ground truth) and Top-5 accuracy (ground truth among the five most likely predictions). As model outputs are unconstrained free text, GPT-4o is used to perform semantic matching between predictions and ground truth labels. The task demands multimodal reasoning and open-ended prediction and is performed in a zero-shot setting, closely mirroring real-world clinical decision-making workflows.
\section{Experiments and Results}

We benchmarked large vision-language models on NOVA to systematically test their ability to generalize under extreme clinical heterogeneity. All experiments were conducted in inference-only mode. We report results for Gemini 2.0 Flash (Google DeepMind), Qwen2-VL-72B (Alibaba DAMO Academy), and Qwen2.5-VL-72B for abnormality grounding; and GPT-4o (OpenAI), Gemini 2.0 Flash, and Qwen2.5-VL-72B-Instruct for image captioning and diagnostic reasoning. As these models are proprietary and their training data is undisclosed, Eurorad cases may have been partially included. Results should thus be interpreted as an upper bound on zero-shot generalization. We encourage future evaluation of open models for a more conservative assessment.

We designed NOVA to expose generalization failures in models confronted with previously unseen, rare clinical cases. To do so, we evaluate along three critical axes of clinical reasoning: \emph{localization, description, and diagnosis.}

\subsection{Stress Test 1: Localization under Clinical Heterogeneity}

\begin{figure}[tb!]
\centering
\includegraphics[width=\linewidth]{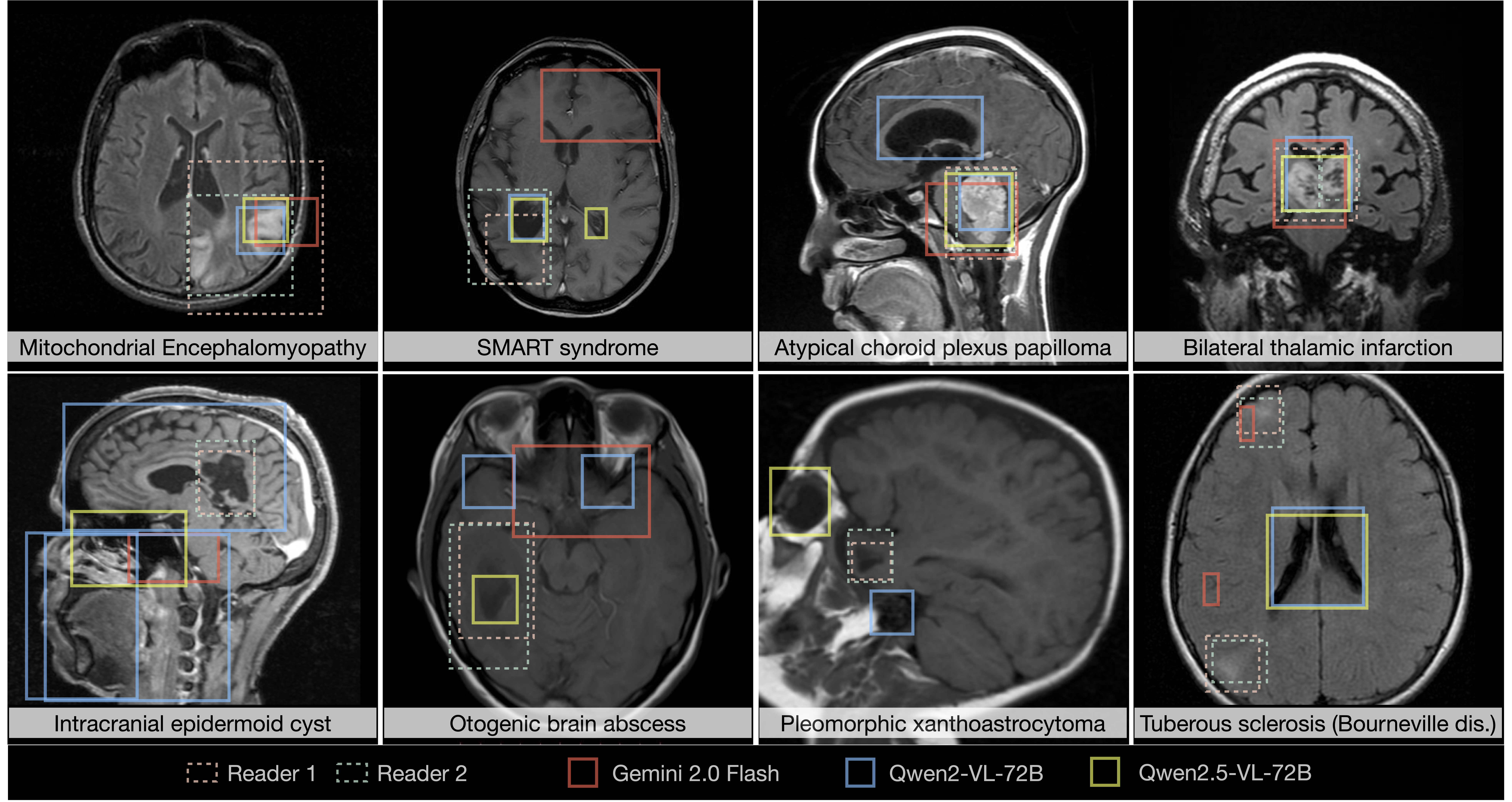}
\caption{Examples of model predictions for anomaly grounding on NOVA. Ground truth and model-predicted bounding boxes are shown for Gemini 2.0 Flash, Qwen2.0-VL-72B, and Qwen2.5-VL-72B.}

\label{fig::detection}
\end{figure}

The anomaly localization task revealed a strong performance degradation. While large vision-language models achieve detection scores of 73\%–92\%~\cite{yang2024qwen2} on natural image benchmarks such as RefCOCO~\cite{yu2016modeling}, performance on NOVA dropped sharply to 8.3\%–28.5\%. Models were evaluated using standard object detection metrics (mAP@30, mAP@50, and mAP@[50:95]), as summarized in Table~\ref{tab::localization}. Despite occasional correct detections, all models exhibited poor calibration under clinical distribution shift, frequently producing incomplete, misplaced, or spurious bounding boxes. Clinical inspection of representative cases (Figure~\ref{fig::detection}) revealed typical failure patterns such as false-positive detection of normal anatomical structures (e.g., orbital cavity misinterpreted as lesion by Qwen2.5-VL) and failure to localize true abnormalities even under relaxed overlap thresholds. Quantitatively, even at 30\% IoU criteria, fewer than half of ground truth abnormalities were detected, and over 600 false-positive boxes were recorded. In clinical practice, missed anomalies risk delayed or missed diagnoses, while high false-positive rates drive unnecessary specialist referrals, patient anxiety, and increased healthcare costs. NOVA sets the first extreme benchmark designed to systematically expose these critical failure modes.

\subsection{Stress Test 2: Description under Semantic Shift}

The second test probes whether models can generate clinically meaningful image descriptions under severe semantic shifts. Table~\ref{tab::captioning_reasoning} summarizes performance across Clinical Term F1, Modality Term F1, BLEU, METEOR, and binary abnormality classification accuracy. Gemini 2.0 Flash achieved the highest Clinical Term F1 (19.8\%), Modality Term F1 (59.8\%), and BLEU (1.83), reflecting comparatively stronger recognition of structured imaging attributes. GPT-4o outperforms in Binary F1 (11.3\%) and METEOR (17.5), indicating slightly better fluency and sentence-level description. Qwen2.5-VL-72B-Instruct underperformed across all metrics.
\begin{table}[tb!]
\centering
\caption{Localization performance on NOVA. We evaluate the models with standard object detection metrics (mAP at multiple thresholds), detection accuracy (ACC50), number of true positives (TP30), and number of false positives (FP30).}

\label{tab::localization}
\adjustbox{max width=\linewidth}{
\begin{tabular}{lccccccc}
\toprule
Model & mAP30 & mAP50 & mAP50-95 & ACC50 & TP30 & FP30 \\ 
\midrule
Gemini 2.0 Flash & 20.16 & 7.37 & 1.99 & 8.83 & 227/1068 & 899 \\ 
Qwen2-VL-72B & 25.02 & 15.09 & 6.44 & 25.50 & 338/1068 & 1163\\ 
Qwen2.5-VL-72B & \textbf{37.66} & \textbf{24.49} & \textbf{11.23} & \textbf{28.48} & \textbf{406/1068} & \textbf{672} \\ 
\bottomrule
\end{tabular}
}
\end{table}

\begin{figure}[tb!]
\centering
\includegraphics[width=\linewidth]{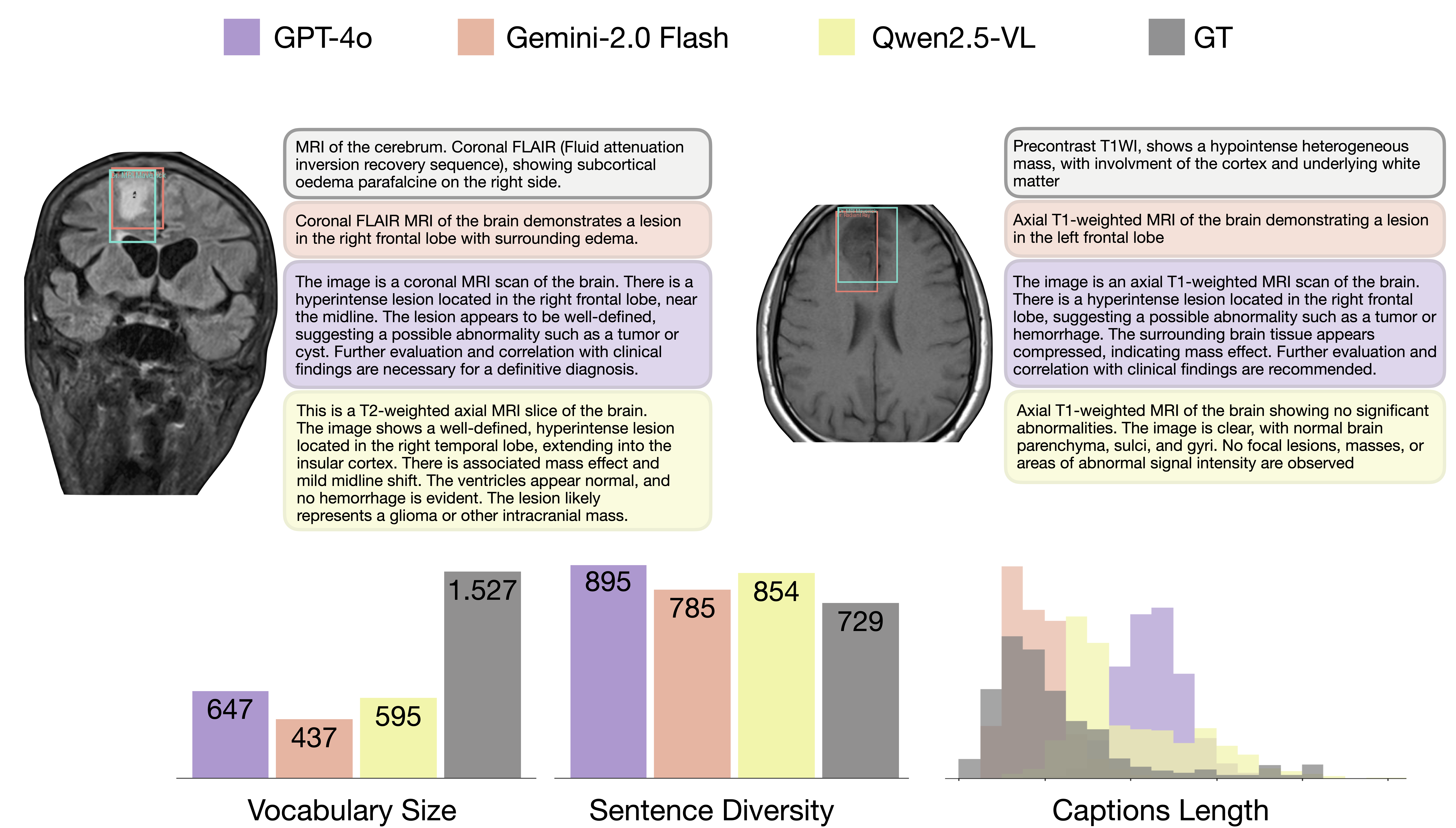}
\caption{Task2: Image Captioning. (Top) Example image-caption pairs showing model predictions and reference ground truth. Model outputs tend toward verbose, redundant phrasing with fewer specialized terms. (Bottom) Quantitative analysis. Left: Vocabulary size (unique words). Center: Sentence diversity (unique captions). Right: Caption length distribution (number of words per caption).
Ground truth radiology reports exhibit the highest vocabulary richness but shorter, information-dense sentences. All models display severe vocabulary collapse and compensate with longer and more varied sentence constructions.}
\label{fig::captions}
\end{figure}

To further investigate the semantic limitations observed in captioning (Table~\ref{tab::captioning_reasoning}), we analyzed language behavior across models in Figure~\ref{fig::captions}. Example cases illustrate that all models tend to produce longer but vaguer descriptions under uncertainty.
The ground truth showed the highest vocabulary size (1527 unique words), reflecting expert use of precise diagnostic terminology. VLMs exhibited drastic vocabulary compression, with GPT-4o, Gemini 2.0 Flash, and Qwen2.5-VL using only 647, 437, and 595 unique words, respectively.
Interestingly, models showed comparable or slightly higher sentence diversity (895, 785, and 854 unique captions vs. 729 in the ground truth), likely due to paraphrasing and verbose redundancy.
Caption length distribution revealed distinct patterns: Gemini produced captions with similar lengths to ground truth, while GPT-4o and Qwen2.5-VL consistently generated longer outputs.
This analysis highlights a consistent pattern of low lexical precision and repetitive verbosity, particularly for GPT-4o and Qwen2.5-VL, confirming that current models exhibit limitations in expressing clinically meaningful descriptive detail in image captioning tasks under real-world clinical distribution shifts.

\subsection{Stress Test 3: Diagnostic Reasoning under Distributional Shift}
\begin{figure}[tb!]
\centering
\begin{minipage}{0.45\linewidth}
\centering
\setlength{\tabcolsep}{2pt} 
\captionof{table}{Diagnostic reasoning results on NOVA. Diagnostic accuracy is captured by the Top-1 and Top-5 accuracy. Coverage and entropy are extracted from diagnostic reasoning distributions.}
\label{tab::reasoning}
\adjustbox{max width=\linewidth}{
\begin{tabular}{lcccc}
\toprule
Model & Top-1 & Top-5 & Cov. & Ent. \\
\midrule
Gemini 2.0 Flash & 22.1 & 37.4 & 29.4 & \textbf{7.71} \\
GPT-4o & \textbf{24.2} & \textbf{38.4} & \textbf{31.9} & 7.64 \\
Qwen2.5-VL-72B & 22.4 & 35.2 & 26.1 & 7.26 \\
\bottomrule
\end{tabular}
}
\end{minipage}%
\hfill
\begin{minipage}{0.54\linewidth}
    \centering
    \includegraphics[width=\linewidth]{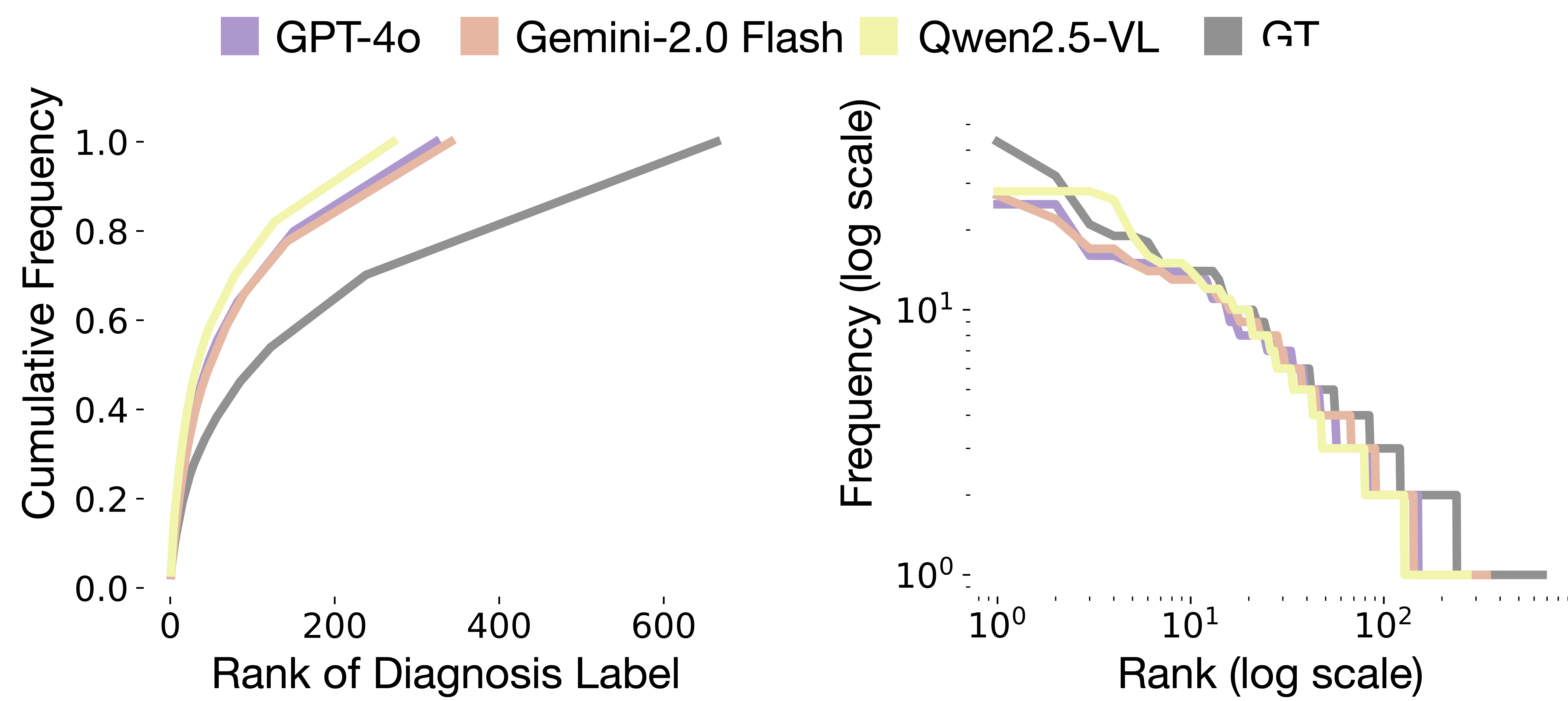}
    \captionof{figure}{Distribution of diagnostic label frequencies for ground truth vs model predictions.}
    \label{fig::reasoning}
\end{minipage}
\end{figure}

The final test evaluates models' ability to assign diagnostic labels based on combined image captions and clinical history. Performance was assessed via Top-1 and Top-5 classification accuracy (Table~\ref{tab::reasoning}). GPT-4o achieved the highest scores (24.2\% Top-1, 38.4\% Top-5), with Gemini 2.0 Flash and Qwen2.5-VL-72B performing lower.

To further probe diagnostic behavior, we analyzed prediction distributions against ground truth (Figure~\ref{fig::reasoning}). All models followed the expected Zipfian-like scaling of disease frequencies (right), demonstrating comparable rank-frequency slopes to ground truth. However, this occurred over a substantially compressed label space, with model predictions collapsing onto smaller vocabularies covering only $\sim$30\% of the ground truth (Table~\ref{tab::reasoning}). This truncation effect was also visible in the cumulative frequency curves (left), where model predictions saturated rapidly relative to ground truth.

Ground truth labels exhibited a Shannon entropy of 8.68 bits, reflecting the high uncertainty and diversity of rare disease distributions. In contrast, model outputs showed a marked entropy reduction of approximately 1 bit (Table~\ref{tab::reasoning}), consistent with over-reliance on dominant classes and poor exploration of the long tail—an effect akin to premature entropy collapse under distributional shift.

NOVA provides the first large-scale benchmark to systematically diagnose and quantify failure modes, and to encourage the development of more robust vision-language reasoning under clinical distribution shift.

\begin{table}[tb!]
\centering
\caption{Image Description on NOVA. Captioning quality is evaluated by Clinical Term F1, Modality Term F1, BLEU, and METEOR. Binary F1 measures binary abnormality classification performance.}
\label{tab::captioning_reasoning}
\adjustbox{max width=\linewidth}{
\begin{tabular}{lccccc}
\toprule
Model & Clinical F1 (\%) & Modality F1 (\%) & BLEU & METEOR & Binary F1 (\%) \\
\midrule
Gemini 2.0 Flash & \textbf{19.8} & \textbf{59.8} & \textbf{1.83} & 15.2 & 5.3  \\
GPT-4o & 15.7 & 49.3 & 0.92 & \textbf{17.5} & \textbf{11.3}  \\
Qwen2.5-VL-72B & 13.6 & 45.3 & 1.08 & 17.1 & 2.4  \\
\bottomrule
\end{tabular}
}
\end{table}

\section{Discussion}

NOVA introduces a new benchmark paradigm for evaluating anomaly detection and multimodal reasoning in clinical brain MRI. Its design offers several key advantages. First, NOVA provides one of the largest and most diverse expert-annotated collections of brain MRI scans available, covering approximately 900 scans with over 280 distinct diagnoses and rare pathological conditions. Second, the dataset uniquely integrates multimodal annotations, combining radiologist-drawn bounding boxes, expert-generated image captions, and detailed clinical histories. This comprehensive structure enables systematic evaluation of detection, description, and diagnostic reasoning within a single resource. Third, the dataset reflects real-world clinical variability by using actual patient cases rather than synthetic perturbations, creating a challenging and realistic testbed. Finally, the structured multi-reader annotation protocol with adjudication by a senior neuroradiologist ensures a high level of annotation quality and reliability.

Despite these advantages, NOVA has limitations that are important to acknowledge. The dataset is sourced from a European radiology teaching repository, which may introduce geographic or demographic biases that could affect model generalization in other healthcare systems. Additionally, NOVA provides only 2D image slices rather than full 3D volumes. 
While this choice may constrain certain volumetric analyses, the decision to release data in 2D format was deliberate: most standard machine learning and computer vision tools and libraries offer limited support for 3D medical imaging, which can significantly slow down experimentation and accessibility for the broader research community.
Finally, NOVA is released as an \emph{evaluation-only} benchmark, not intended for supervised model training. This design reflects the realities of rare disease imaging, where collecting sufficient labeled data for training is often infeasible and where true generalization must be tested without model adaptation.

Looking forward, we plan to maintain NOVA as a dynamic benchmark and to open a public leaderboard to encourage continuous community participation and advancement of the state of the art. Given the dataset's focus on rare diseases and its intended role as an inference benchmark, we do not envision extensions to fine-tuning tasks or inclusion of 3D imaging data. Instead, we anticipate that NOVA will catalyze the development of next-generation foundation models and vision-language systems capable of performing robust anomaly detection and clinical reasoning under realistic open-set clinical conditions.
\section{Conclusion}
We present NOVA, the first large-scale, expert-annotated benchmark dataset for anomaly localization, clinical captioning, and diagnostic reasoning in brain MRI. NOVA provides a uniquely challenging and clinically grounded resource, combining real-world imaging variability with high-quality multimodal annotations. By releasing NOVA to the community, we aim to establish a new standard for evaluating the robustness and generalization of models for clinical anomaly detection and multimodal medical reasoning. We invite the research community to engage with NOVA and drive the development of next-generation models capable of detecting the unknown in clinical imaging.
\bibliographystyle{plain}
\bibliography{neurips_2025}

\begin{thebibliography}{10}

\bibitem{ixi}
Ixi dataset.
\newblock \url{https://brain-development.org/ixi-dataset/}.
\newblock Accessed: 2023-02-15.

\bibitem{banerjee-lavie-2005-meteor}
Satanjeev Banerjee and Alon Lavie.
\newblock {METEOR}: An automatic metric for {MT} evaluation with improved
  correlation with human judgments.
\newblock In Jade Goldstein, Alon Lavie, Chin-Yew Lin, and Clare Voss, editors,
  {\em Proceedings of the {ACL} Workshop on Intrinsic and Extrinsic Evaluation
  Measures for Machine Translation and/or Summarization}, pages 65--72, Ann
  Arbor, Michigan, June 2005. Association for Computational Linguistics.

\bibitem{bercea2023generalizing}
Cosmin~I. Bercea, Benedikt Wiestler, Daniel Rueckert, and Julia~A Schnabel.
\newblock Generalizing unsupervised anomaly detection: Towards unbiased
  pathology screening.
\newblock In {\em Medical Imaging with Deep Learning}, 2023.

\bibitem{bercea2024diffusion}
Cosmin~I Bercea, Benedikt Wiestler, Daniel Rueckert, and Julia~A Schnabel.
\newblock Diffusion models with implicit guidance for medical anomaly
  detection.
\newblock In {\em International Conference on Medical Image Computing and
  Computer-Assisted Intervention}, pages 211--220. Springer, 2024.

\bibitem{bercea2025evaluating}
Cosmin~I Bercea, Benedikt Wiestler, Daniel Rueckert, and Julia~A Schnabel.
\newblock Evaluating normative representation learning in generative ai for
  robust anomaly detection in brain imaging.
\newblock {\em Nature Communications}, 16(1):1624, 2025.

\bibitem{bergmann2019mvtec}
Paul Bergmann, Michael Fauser, David Sattlegger, and Carsten Steger.
\newblock Mvtec ad — a comprehensive real-world dataset for unsupervised
  anomaly detection.
\newblock In {\em 2019 IEEE/CVF Conference on Computer Vision and Pattern
  Recognition (CVPR)}, pages 9584--9592, 2019.

\bibitem{padchest}
Agust{\'i}n Bustos, Antonio Pertusa, Jose~M. Salinas, and Mar{\'i}a de~la
  Iglesia-Vay{\'a}.
\newblock Padchest: A large chest x-ray image dataset with multi-label
  annotated reports.
\newblock {\em Medical Image Analysis}, 66:101797, 2020.

\bibitem{padchest-gr}
Daniel~C Castro, Aurelia Bustos, Shruthi Bannur, Stephanie~L Hyland, Kenza
  Bouzid, Maria~Teodora Wetscherek, Maria~Dolores S{\'a}nchez-Valverde, Lara
  Jaques-P{\'e}rez, Lourdes P{\'e}rez-Rodr{\'\i}guez, Kenji Takeda, et~al.
\newblock Padchest-gr: A bilingual chest x-ray dataset for grounded radiology
  report generation.
\newblock {\em arXiv preprint arXiv:2411.05085}, 2024.

\bibitem{chen2020unsupervised}
Xiaoran Chen, Suhang You, Kerem~Can Tezcan, and Ender Konukoglu.
\newblock Unsupervised lesion detection via image restoration with a normative
  prior.
\newblock {\em Medical Image Analysis}, 64:101713, 2020.

\bibitem{dosovitskiy2020image}
Alexey Dosovitskiy, Lucas Beyer, Alexander Kolesnikov, Dirk Weissenborn,
  Xiaohua Zhai, Thomas Unterthiner, Mostafa Dehghani, Matthias Minderer, Georg
  Heigold, Sylvain Gelly, et~al.
\newblock An image is worth 16x16 words: Transformers for image recognition at
  scale.
\newblock {\em arXiv preprint arXiv:2010.11929}, 2020.

\bibitem{feng2022promptdet}
Chengjian Feng, Yujie Zhong, Zequn Jie, Xiangxiang Chu, Haibing Ren, Xiaolin
  Wei, Weidi Xie, and Lin Ma.
\newblock Promptdet: Towards open-vocabulary detection using uncurated images.
\newblock In {\em European conference on computer vision}, pages 701--717.
  Springer, 2022.

\bibitem{ood3}
Jia Guo, Shuai Lu, Lize Jia, Weihang Zhang, and Huiqi Li.
\newblock Recontrast: Domain-specific anomaly detection via contrastive
  reconstruction.
\newblock In A.~Oh, T.~Naumann, A.~Globerson, K.~Saenko, M.~Hardt, and
  S.~Levine, editors, {\em Advances in Neural Information Processing Systems},
  volume~36, pages 10721--10740, 2023.

\bibitem{hamamci2024developing}
Ibrahim~Ethem Hamamci, Sezgin Er, Furkan Almas, et~al.
\newblock Developing generalist foundation models from a multimodal dataset for
  3d computed tomography.
\newblock {\em arXiv preprint arXiv:2403.17834}, 2024.

\bibitem{hendrycks2018deep}
Dan Hendrycks, Mantas Mazeika, and Thomas Dietterich.
\newblock Deep anomaly detection with outlier exposure.
\newblock {\em arXiv preprint arXiv:1812.04606}, 2018.

\bibitem{isles}
Moritz~R Hernandez~Petzsche, Ezequiel de~la Rosa, Uta Hanning, Roland Wiest,
  Waldo Valenzuela, Mauricio Reyes, Maria Meyer, Sook-Lei Liew, Florian Kofler,
  Ivan Ezhov, et~al.
\newblock Isles 2022: A multi-center magnetic resonance imaging stroke lesion
  segmentation dataset.
\newblock {\em Scientific Data}, 9(1):762, 2022.

\bibitem{hong2024out}
Zesheng Hong, Yubiao Yue, Yubin Chen, Lele Cong, Huanjie Lin, Yuanmei Luo,
  Mini~Han Wang, Weidong Wang, Jialong Xu, Xiaoqi Yang, et~al.
\newblock Out-of-distribution detection in medical image analysis: A survey.
\newblock {\em arXiv preprint arXiv:2404.18279}, 2024.

\bibitem{mimic}
Alistair E.~W. Johnson, Tom~J. Pollard, Seth~J. Berkowitz, et~al.
\newblock {MIMIC-CXR}, a de-identified publicly available database of chest
  radiographs with free-text reports.
\newblock {\em Scientific Data}, 6:317, 2019.

\bibitem{Kim2014Fool}
Y.~W. Kim and L.~T. Mansfield.
\newblock Fool me twice: Delayed diagnoses in radiology with emphasis on
  perpetuated errors.
\newblock {\em AJR. American Journal of Roentgenology}, 202(3):465--470, 2014.

\bibitem{koh2021wilds}
Pang~Wei Koh, Shiori Sagawa, Henrik Marklund, Sang~Michael Xie, Marvin Zhang,
  Akshay Balsubramani, Weihua Hu, Michihiro Yasunaga, Richard~Lanas Phillips,
  Irena Gao, Tony Lee, Etienne David, Ian Stavness, Wei Guo, Berton Earnshaw,
  Imran Haque, Sara~M Beery, Jure Leskovec, Anshul Kundaje, Emma Pierson,
  Sergey Levine, Chelsea Finn, and Percy Liang.
\newblock Wilds: A benchmark of in-the-wild distribution shifts.
\newblock In Marina Meila and Tong Zhang, editors, {\em Proceedings of the 38th
  International Conference on Machine Learning}, volume 139 of {\em Proceedings
  of Machine Learning Research}, pages 5637--5664. PMLR, 18--24 Jul 2021.

\bibitem{emergency}
Seungjun Lee, Boryeong Jeong, Minjee Kim, Ryoungwoo Jang, Wooyul Paik, Jiseon
  Kang, Won Chung, Gil-Sun Hong, and Namkug Kim.
\newblock Emergency triage of brain computed tomography via anomaly detection
  with a deep generative model.
\newblock {\em Nature Communications}, 13:4251, 07 2022.

\bibitem{mslub}
{\v{Z}}iga Lesjak, Alina Galimzianova, Andrej Koren, et~al.
\newblock A novel public {MR} image dataset of multiple sclerosis patients with
  lesion segmentations based on multi-rater consensus.
\newblock {\em Neuroinformatics}, 16(1):51--63, 2018.

\bibitem{li2021cutpaste}
Chun-Liang Li, Kihyuk Sohn, Jinsung Yoon, and Tomas Pfister.
\newblock Cutpaste: Self-supervised learning for anomaly detection and
  localization.
\newblock In {\em Proceedings of the IEEE/CVF conference on computer vision and
  pattern recognition}, pages 9664--9674, 2021.

\bibitem{atlas}
Sook-Lei Liew, Bethany~P. Lo, ., and et~al. Miarnda R.~Donnelly.
\newblock A large, curated, open-source stroke neuroimaging dataset to improve
  lesion segmentation algorithms.
\newblock {\em Scientific Data}, 9, 2022.

\bibitem{mahmood2025benchmarking}
Faisal Mahmood.
\newblock A benchmarking crisis in biomedical machine learning.
\newblock {\em Nature Medicine}, 31(4):1060--1060, 2025.

\bibitem{oasis}
Daniel~S. Marcus, Aditya~F. Fotenos, John~G. Csernansky, John~C. Morris, and
  Randy~L. Buckner.
\newblock Open access series of imaging studies: longitudinal {MRI} data in
  nondemented and demented older adults.
\newblock {\em Journal of Cognitive Neuroscience}, 22(12):2677--2684, 2010.

\bibitem{menze2015brats}
Bjoern~H. Menze, Andras Jakab, Stefan Bauer, Jayashree Kalpathy-Cramer, Keyvan
  Farahani, Justin Kirby, Yuliya Burren, Nicole Porz, Johannes Slotboom, Roland
  Wiest, Levente Lanczi, Elizabeth Gerstner, Marc-André Weber, Tal Arbel,
  Brian~B. Avants, Nicholas Ayache, Patricia Buendia, D.~Louis Collins, Nicolas
  Cordier, Jason~J. Corso, Antonio Criminisi, Tilak Das, Hervé Delingette,
  Çağatay Demiralp, Christopher~R. Durst, Michel Dojat, Senan Doyle, Joana
  Festa, Florence Forbes, Ezequiel Geremia, Ben Glocker, Polina Golland,
  Xiaotao Guo, Andac Hamamci, Khan~M. Iftekharuddin, Raj Jena, Nigel~M. John,
  Ender Konukoglu, Danial Lashkari, José~António Mariz, Raphael Meier,
  Sérgio Pereira, Doina Precup, Stephen~J. Price, Tammy~Riklin Raviv, Syed
  M.~S. Reza, Michael Ryan, Duygu Sarikaya, Lawrence Schwartz, Hoo-Chang Shin,
  Jamie Shotton, Carlos~A. Silva, Nuno Sousa, Nagesh~K. Subbanna, Gabor
  Szekely, Thomas~J. Taylor, Owen~M. Thomas, Nicholas~J. Tustison, Gozde Unal,
  Flor Vasseur, Max Wintermark, Dong~Hye Ye, Liang Zhao, Binsheng Zhao, Darko
  Zikic, Marcel Prastawa, Mauricio Reyes, and Koen Van~Leemput.
\newblock The multimodal brain tumor image segmentation benchmark (brats).
\newblock {\em IEEE Transactions on Medical Imaging}, 34(10):1993--2024, 2015.

\bibitem{papineni-etal-2002-bleu}
Kishore Papineni, Salim Roukos, Todd Ward, and Wei-Jing Zhu.
\newblock {B}leu: a method for automatic evaluation of machine translation.
\newblock In Pierre Isabelle, Eugene Charniak, and Dekang Lin, editors, {\em
  Proceedings of the 40th Annual Meeting of the Association for Computational
  Linguistics}, pages 311--318, Philadelphia, Pennsylvania, USA, July 2002.
  Association for Computational Linguistics.

\bibitem{adni}
Ronald~C. Petersen, Paul~S. Aisen, Laurel~A. Beckett, Michael~C. Donohue,
  Anthony~C. Gamst, Danielle~J. Harvey, Clifford R.~Jr. Jack, William~J.
  Jagust, Leslie~M. Shaw, Arthur~W. Toga, John~Q. Trojanowski, and Michael~W.
  Weiner.
\newblock Alzheimer's disease neuroimaging initiative (adni): clinical
  characterization.
\newblock {\em Neurology}, 74(3):201--209, January 2010.

\bibitem{pidhorskyi2018generative}
Stanislav Pidhorskyi, Ranya Almohsen, and Gianfranco Doretto.
\newblock Generative probabilistic novelty detection with adversarial
  autoencoders.
\newblock {\em Advances in neural information processing systems}, 31, 2018.

\bibitem{radford2021learning}
Alec Radford, Jong~Wook Kim, Chris Hallacy, Aditya Ramesh, Gabriel Goh,
  Sandhini Agarwal, Girish Sastry, Amanda Askell, Pamela Mishkin, Jack Clark,
  et~al.
\newblock Learning transferable visual models from natural language
  supervision.
\newblock In {\em International Conference on Machine Learning}, pages
  8748--8763. PmLR, 2021.

\bibitem{recht2019imagenet}
Benjamin Recht, Rebecca Roelofs, Ludwig Schmidt, and Vaishaal Shankar.
\newblock Do imagenet classifiers generalize to imagenet?
\newblock In {\em Proceedings of the International Conference on Machine
  Learning}, pages 5389--5400. PMLR, 2019.

\bibitem{brain}
Carola Reinhard, Anne-Catherine Bachoud-Lévi, Tobias Bäumer, Enrico Bertini,
  Alicia Brunelle, Annemieke~I. Buizer, Antonio Federico, Thomas Gasser, Samuel
  Groeschel, Sanja Hermanns, Thomas Klockgether, Ingeborg Krägeloh-Mann,
  G.~Bernhard Landwehrmeyer, Isabelle Leber, Alfons Macaya, Caterina Mariotti,
  Wassilios~G. Meissner, Maria~Judit Molnar, Jorik Nonnekes, Juan~Dario
  Ortigoza~Escobar, Belen Pérez~Dueñas, Lori Renna~Linton, Ludger Schöls,
  Rebecca Schuele, Marina A.~J. Tijssen, Rik Vandenberghe, Anna Volkmer,
  Nicole~I. Wolf, and Holm Graessner.
\newblock The european reference network for rare neurological diseases.
\newblock {\em Frontiers in Neurology}, Volume 11 - 2020, 2021.

\bibitem{roth2022towards}
Karsten Roth, Latha Pemula, Joaquin Zepeda, Bernhard Sch{\"o}lkopf, Thomas
  Brox, and Peter Gehler.
\newblock Towards total recall in industrial anomaly detection.
\newblock In {\em Proceedings of the IEEE/CVF conference on computer vision and
  pattern recognition}, pages 14318--14328, 2022.

\bibitem{ruff2021unifying}
Lukas Ruff, Jacob~R Kauffmann, Robert~A Vandermeulen, Gr{\'e}goire Montavon,
  Wojciech Samek, Marius Kloft, Thomas~G Dietterich, and Klaus-Robert
  M{\"u}ller.
\newblock A unifying review of deep and shallow anomaly detection.
\newblock {\em Proceedings of the IEEE}, 109(5):756--795, 2021.

\bibitem{schlegl2019f}
Thomas Schlegl, Philipp Seeb{\"o}ck, Sebastian~M Waldstein, Georg Langs, and
  Ursula Schmidt-Erfurth.
\newblock f-anogan: Fast unsupervised anomaly detection with generative
  adversarial networks.
\newblock {\em Medical Image Analysis}, 54:30--44, 2019.

\bibitem{schluter2022natural}
Hannah~M Schl{\"u}ter, Jeremy Tan, Benjamin Hou, and Bernhard Kainz.
\newblock Natural synthetic anomalies for self-supervised anomaly detection and
  localization.
\newblock In {\em European Conference on Computer Vision}, pages 474--489.
  Springer, 2022.

\bibitem{ukbiobank}
Cathie Sudlow, John Gallacher, Naomi Allen, Valerie Beral, Paul Burton, John
  Danesh, et~al.
\newblock Uk biobank: An open access resource for identifying the causes of a
  wide range of complex diseases of middle and old age.
\newblock {\em PLoS Medicine}, 12(3):e1001779, 2015.

\bibitem{camcan}
Jason~R. Taylor, Nitin Williams, Rhodri Cusack, Tibor Auer, Meredith~A. Shafto,
  Marie Dixon, Lorraine~K. Tyler, Cam-CAN, and Richard~N. Henson.
\newblock The cambridge centre for ageing and neuroscience (cam-can) data
  repository: Structural and functional mri, meg, and cognitive data from a
  cross-sectional adult lifespan sample.
\newblock {\em NeuroImage}, 144:262--269, 2017.
\newblock Data Sharing Part II.

\bibitem{wang2024bridging}
Han Wang and Yixuan Li.
\newblock Bridging ood detection and generalization: A graph-theoretic view.
\newblock {\em Advances in Neural Information Processing Systems}, 2024.

\bibitem{wolleb2022diffusion}
Julia Wolleb, Florentin Bieder, Robin Sandk{\"u}hler, and Philippe~C Cattin.
\newblock Diffusion models for medical anomaly detection.
\newblock In {\em International Conference on Medical Image Computing and
  Computer-Assisted Intervention}, pages 35--45. Springer, 2022.

\bibitem{yang2024qwen2}
An~Yang, Baosong Yang, Beichen Zhang, Binyuan Hui, Bo~Zheng, Bowen Yu,
  Chengyuan Li, Dayiheng Liu, Fei Huang, Haoran Wei, et~al.
\newblock Qwen2. 5 technical report.
\newblock {\em arXiv preprint arXiv:2412.15115}, 2024.

\bibitem{ood2}
Jingkang Yang, Pengyun Wang, Dejian Zou, Zitang Zhou, Kunyuan Ding, WENXUAN
  PENG, Haoqi Wang, Guangyao Chen, Bo~Li, Yiyou Sun, Xuefeng Du, Kaiyang Zhou,
  Wayne Zhang, Dan Hendrycks, Yixuan Li, and Ziwei Liu.
\newblock Openood: Benchmarking generalized out-of-distribution detection.
\newblock In S.~Koyejo, S.~Mohamed, A.~Agarwal, D.~Belgrave, K.~Cho, and A.~Oh,
  editors, {\em Advances in Neural Information Processing Systems}, volume~35,
  pages 32598--32611, 2022.

\bibitem{yu2016modeling}
Licheng Yu, Patrick Poirson, Shan Yang, Alexander~C Berg, and Tamara~L Berg.
\newblock Modeling context in referring expressions.
\newblock In {\em European Conference on Computer Vision}, pages 69--85.
  Springer, 2016.

\bibitem{zbontar2018fastmri}
Jure Zbontar, Florian Knoll, Anuroop Sriram, Tullie Murrell, Zhengnan Huang,
  Matthew~J Muckley, Aaron Defazio, Ruben Stern, Patricia Johnson, Mary Bruno,
  et~al.
\newblock fastmri: An open dataset and benchmarks for accelerated mri.
\newblock {\em arXiv preprint arXiv:1811.08839}, 2018.

\bibitem{zhao2022fastmri+}
Ruiyang Zhao, Burhaneddin Yaman, Yuxin Zhang, Russell Stewart, Austin Dixon,
  Florian Knoll, Zhengnan Huang, Yvonne~W Lui, Michael~S Hansen, and Matthew~P
  Lungren.
\newblock fastmri+, clinical pathology annotations for knee and brain fully
  sampled magnetic resonance imaging data.
\newblock {\em Scientific Data}, 9(1):152, 2022.

\bibitem{ood1}
Haotian Zheng, Qizhou Wang, Zhen Fang, Xiaobo Xia, Feng Liu, Tongliang Liu, and
  Bo~Han.
\newblock Out-of-distribution detection learning with unreliable
  out-of-distribution sources.
\newblock In A.~Oh, T.~Naumann, A.~Globerson, K.~Saenko, M.~Hardt, and
  S.~Levine, editors, {\em Advances in Neural Information Processing Systems},
  volume~36, pages 72110--72123, 2023.

\bibitem{zhu2024survey}
Chaoyang Zhu and Long Chen.
\newblock A survey on open-vocabulary detection and segmentation: Past,
  present, and future.
\newblock {\em IEEE Transactions on Pattern Analysis and Machine Intelligence},
  2024.

\bibitem{zimmerer2022mood}
David Zimmerer, Peter~M Full, Fabian Isensee, Paul J{\"a}ger, Tim Adler, Jens
  Petersen, Gregor K{\"o}hler, Tobias Ross, Annika Reinke, Antanas Kascenas,
  et~al.
\newblock Mood 2020: A public benchmark for out-of-distribution detection and
  localization on medical images.
\newblock {\em IEEE Transactions on Medical Imaging}, 41(10):2728--2738, 2022.

\bibitem{zimmerer2019unsupervised}
David Zimmerer, Fabian Isensee, Jens Petersen, Simon Kohl, and Klaus
  Maier-Hein.
\newblock Unsupervised anomaly localization using variational auto-encoders.
\newblock In {\em Medical Image Computing and Computer Assisted
  Intervention--MICCAI 2019: 22nd International Conference, Shenzhen, China,
  October 13--17, 2019, Proceedings, Part IV 22}, pages 289--297. Springer,
  2019.

\end{thebibliography}


\begin{ack}
C.I.B. is funded via the EVUK program (”Next-generation AI for Integrated Diagnostics”) of the Free State of Bavaria.
\end{ack}

\end{document}